\documentclass{iau}
\usepackage{graphicx}
\usepackage{bm}

\title[Preliminary design of the INPE's Solar Vector Magnetograph] 
{Preliminary design of the INPE's Solar Vector Magnetograph}

\author[INPE's Solar Vector Magnetograph Team]   
{L.~E.~A. Vieira$^{1}$, A.~L. Cl\'ua de Gonzalez$^{1}$, A.~Dal Lago$^{1}$, C.~Wrasse$^{1}$,E.~Echer$^{1}$,F.L.Guarnieri$^{1}$,  F.~Reis~Cardoso$^{2}$, G.~Guerrero$^{3}$, J.~Rezende Costa$^{1}$, J.~Palacios$^{4}$, L.~Balmaceda$^{1,5}$, L.~Ribeiro Alves$^{1}$, L.~da Silva$^{1}$, L.L. Costa$^{1}$, M.~Sampaio$^{1}$, M.~C. Rabello Soares$^{3}$, M.~Barbosa$^{1}$, M.~Domingues$^{1}$, N.~Rigozo$^{1}$, O.~Mendes Jr.$^{1}$, P.~Jauer$^{1}$, R.~Dallaqua$^{1}$, R.H. Branco$^{1}$, T.~Stekel$^{1}$, W.~Gonzalez$^{1}$, W.~Kabata$^{1}$
}

\affiliation{$^1$ Instituto Nacional de Pesquisas Espaciais (INPE),
Avda. dos Astronautas, S\~ao Jos\'e dos Campos--SP,  12227-010, Brazil,
email: {\tt luis.vieira@inpe.br}

$^2$  Universidadede de Sao Paulo, Lorena, SP,  Brazil\\

$^3$ Universidade Federal de Minas Gerais, UFMG-MEC, Belo Horizonte, MG, Brazil\\

$^4$Space Reseach Group--Space Weather, Departamento de F\'isica y Matem\'aticas,  Universidad de Alcal\'a,
University Campus, Sciences Building, P.O. 28871, Alcal\'a de Henares, Spain \\
 
$^5$Instituto de Ciencias Astron\'omicas, de la Tierra y el Espacio, ICATE-CONICET \\ 
Avda. de Espa\~na Sur 1512, J5402DSP, San Juan, Argentina\\
  
}

\pubyear{2015}
\volume{305}  
\jname{Polarimetry: From the Sun to Stars and Stellar Environments}
\editors{K.N. Nagendra, S. Bagnulo, \\ R. Centeno, \& M. Mart\'inez Gonz\'alez, eds.}
\begin{document}

\maketitle

\def\deg{\mbox{$^{\circ}$}}

\begin{abstract}

We describe the preliminary design of a magnetograph and visible-light imager instrument to study the solar dynamo processes through observations of the solar surface magnetic field distribution. The instrument will provide measurements of the vector magnetic field and of the line-of-sight velocity in the solar photosphere. As the magnetic field anchored at the solar surface produces most of the structures and energetic events in the upper solar atmosphere and significantly influences the heliosphere, the development of this instrument plays an important role in reaching the scientific goals of The Atmospheric and Space Science Coordination (CEA) at the Brazilian National Institute for Space Research (INPE). In particular, the CEA's space weather program will benefit most from the development of this technology. We expect that this project will be the starting point to establish a strong research program on Solar Physics in Brazil. Our main aim is acquiring progressively the know-how to build state-of-art solar vector magnetograph and visible-light imagers for space-based platforms to contribute to the efforts of the solar-terrestrial physics community to address the main unanswered questions on how our nearby Star works.


\end{abstract}

\firstsection 
\section{Introduction}
\label{intro}

Living in the surroundings of the atmosphere of a highly variable star allows us to observe in high spatial and temporal resolution the universal processes that occur in its outer layers. This is our main motivation to propose the development of instruments and models to study the evolution of the magnetic structure of the Sun on timescales that range from seconds to millenia. 

The solar electromagnetic and corpuscular emissions are strongly modulated by the evolution of the solar magnetic field. Systematic observations of sunspots, since the invention of the telescope, are the main indicator that the solar activity changes on timescales from days to millennia. These changes drive long-term evolution of the heliosphere (space climate) as well as violent events (space weather). In particular, the near-Earth region is strongly affected by the evolution of the solar magnetic structure.

Early observations of the sunspots clearly indicated that their presence on the solar surface varies cyclically with a period of approximately 11 years, the so-called solar activity cycle. Subsequently, it was observed that the latitudinal distribution of sunspots varies throughout the cycle and follows a pattern that begins at middle latitudes ($\pm 35^\circ$), reaches a maximum, and decays near the equator ($\pm 5^\circ$). Following a minimum of activity, the pattern repeats itself. 

Magnetic field is found at different spatial scales all over the solar surface. Most part of the magnetic flux is filamented into a range of flux tubes with field strength of 100\,mT while the largest photospheric magnetic flux tubes are sunspots with diameters between a few Mm and 50\,Mm, which represent the biggest accumulations of magnetic flux in the photosphere. Additionally, observations indicate that sunspots in the same hemisphere but belonging to distinct cycles have opposite polarities. Consequently, the predominant magnetic field signal in each hemisphere varies with a period of approximately 22 years. 

Dark features appearing on the solar surface (i.e. sunspots and pores) cause a detectable depletion in the flux density. This depletion occurs because intense magnetic fields within sunspots block the convection and decreases the transport of thermal energy from the base of the convection zone to the photosphere. The reduction of the temperature within the sunspots causes a reduction of the surface opacity. Note that the depletion depends on the relative position of the sunspot on the solar surface and the observer. The maximum depletion occurs when the sunspots are near the disk center and can be as large as $\sim$ 0.3\%. On the other hand, small flux tubes (e.g., faculae) appear brighter than the surrounding average solar surface because their partially evacuated interior is efficiently heated by radiation from their surroundings in the deeper layers and, presumably, also by dissipation of mechanical energy in their higher atmospheric layers. Averaged over the whole Sun, the enhanced brightness of the magnetic elements dominates over the reduced energy flux in the sunspots, so that the total radiation output increases with growing magnetic flux in the photosphere. As a consequence, the brightness of the Sun increases by about 0.1\% from minimum to maximum during the solar activity cycle (see e.g. \cite[Domingo et al. 2009]{Domingo2009-vie}, \cite[{Fr{\"o}hlich} 2013]{froehlich2013-vie}).

Long-term changes of the solar activity are observed directly from, among other indices, the sunspot records and indirectly in natural archives such as cosmogenic isotopes. These proxies indicate periods of high (grand maxima) and low (grand minima) solar activity. These changes in the solar energy output clearly impact the Earth's climate, although the relative role of the several drivers (solar, volcanic, anthropogenic, etc) are still under debate. We emphasize that these long-term changes impact not just the troposphere-land-ocean system, but also the neutral and ionized components of the middle and upper atmosphere (\cite[Solanki et al. 2013]{solanki2013-vie}).

In spite of the increasing interest of the solar physics community to unveil the mechanisms responsible for such variations, still there is no complete physical explanation for the origin of the observed solar activity. So far the preferred paradigm used to explain the production and periodic occurrence of sunspots is the action of the magnetohydrodynamic dynamo driven by the differential rotation of the star and the convection of the magnetic field to the surface. At the moment, dynamo models can reproduce cyclic modulation of the Sun's activity, and even some features of the 11-year sunspot cycle, but cannot explain the varying amplitudes of the maxima or long-term changes in the Sun's magnetic activity. 

The magnetic field has also an important role as the driver and energy source for highly dynamical and energetic events such as flares and coronal mass ejections that take place in the outer layers of the solar atmosphere. Measurements of the full vector magnetic field result indispensable in order to answer the main open questions on the origin and evolution of such phenomena.

In this context, we have proposed to develop a magnetograph and visible-light imager instrument in order to study the solar dynamo processes that give rise to the rich variety of phenomena observed at different layers of the solar atmosphere through observations of the surface magnetic field distribution.

\section{Instrument design}

\begin{figure}[t]
\begin{center}
\includegraphics[scale=0.30]{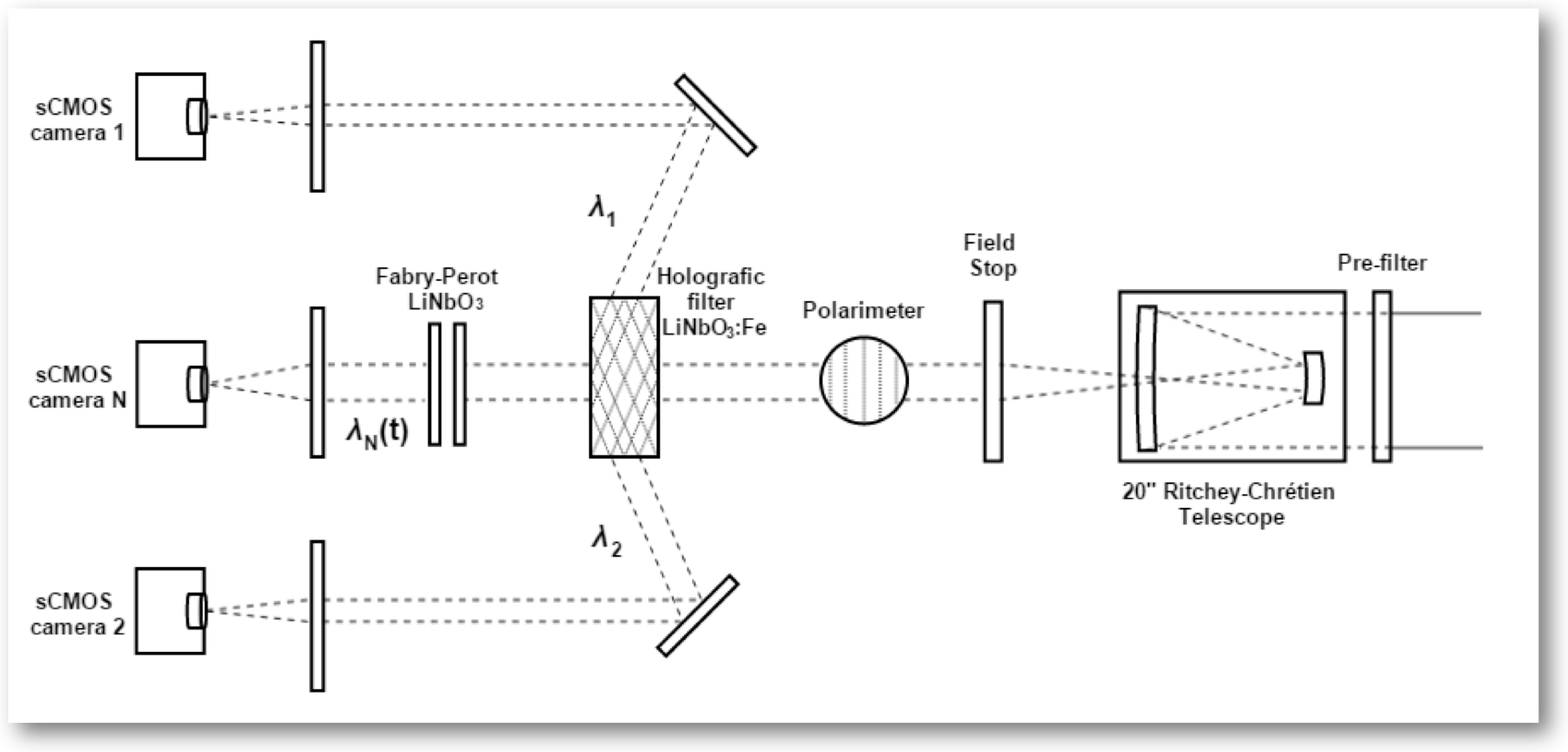}
\caption{Optical design employing a photorefractive crystal with a narrow-band holographic filter. Based on \cite[Keller \& Harvey (1996)]{Keller1996-vie}. }
\label{optical_design}
\end{center}
\end{figure}

Figure~\ref{optical_design} shows a schematic layout of the optical system of the Solar telescope. The instrument will observe the full solar disk in the Fe \textsc{i} absorption lines at 525 nm or 630.2/630.1~nm, as already done by Hinode-SOT/SP (\cite[Suematsu et al. 2008]{suematsu2008-vie}). It consists of a German-equatorial Ritchey-Chr\'etien Telescope of 500 mm aperture and the focal length of 4000 mm (which makes an f/8 beam), a polarization selector, an image stabilization system, a narrowband tunable filter and two sCMOS cameras for orthogonal polarimetry. The specifications of the Ritchey-Chr\'etien telescope are presented in Table~\ref{vieira-tab1}. That configuration may reach a maximum resolution (diffraction limited) of approximately 2\,arcsec, that means without considering the contribution on smearing and blurring of the atmosphere. The field of view (FOV) of the instrument before the polarization module is about 69\,arcmin. 

The light passing through the aperture is analyzed for polarization using a module that consists of two quarter-wave plates mounted on a high resolution rotary stage. These wave-plates, which are compound  zero-order quartz wave-plates, are used to modulate the polarized light signals. The orientation of the wave-plates will be adjusted through a computer control to filter out desired Stokes parameters. A total of four measurement positions are required to measure the complete Stokes vector.  

We also need a polarization calibration unit consisting of a linear polarizer and a quarter wave plate  (and perhaps an additional heat load rejection filter in front) that can be placed in front of the polarization modulator. These elements need to be rotated to various angles either manually or with a motorized mounting.

The collimated beam will passes through a volume holographic grating (VHG) that will allow us to sample simultaneously up to 10 lines. The original concept was proposed by Christoph Keller and Jack Harvey (\cite[Keller \& Harvey 1996]{Keller1996-vie}) in a prospective study for the NASA Solar Probe Mission.  We point out that although VHGs were previously employed as a narrow-band filters for solar observations, magnetographs based on this concept were not yet implemented. Alternatively, the collimated beam will passes through a Lithium Niobate Fabry-P\'erot Etalon. The filter package will be maintained in a temperature controlled box.  

The detector is a crucial item. With ground-based seeing, we need to simultaneously obtain both states of polarization emerging from the modulator and it is highly desirable to run the modulation as fast as possible to reduce seeing noise. In this way, we will employ a new so-called Scientific Complementary Metal-Oxide-Semiconductor (sCMOS) cameras. The cameras have been recently acquired through an international competition. The features and specifications required for the two sCMOS cameras were determined in order to satisfy the specification for the instrument. Specifically, the features required are determined by the spatial resolution, dynamic range, precision photometry and reading noise. 

In an initial phase, we propose to house the entire system on the roof-top of the CEA-II building at INPE's Campus in S\~ao Jos\'e dos Campos, SP, Brazil. The entire rail is proposed to be mounted in a German Equatorial mount.  The individual modules will be fixed on movable plates, which will be then fixed on the optical rail at the desired locations. We are discussing the possibility to install in a second phase the telescope at the Observat\'orio do Pico dos Dias, 1864 m height.   

The control software for the instrument will be developed in house using c and c++ languages. The images from the camera will be saved in FITS (Flexible Image Transport System) format as three-dimensional data cubes and visualized/processed using the MATLAB platform. 

\begin{table}
  \begin{center}
  \caption{Preliminary specifications for the Ritchey-Chr\'etien telescope.}
  \label{vieira-tab1}
 {\scriptsize
  \begin{tabular}{|l|l|}\hline
{Model} & {Ritchey-Chr\'etien} \\ \hline
Aperture & 500 mm  \\ \hline
Focal Length & 4000 mm  \\ \hline
Focal Ratio & f/8  \\ \hline
Back Focus & 365 mm  \\ \hline
Field of View & 69 arc minutes  \\ \hline
Coating & Coating with 96\% Reflection  \\ \hline
Surface quality & Wavefront higher than 95 strehl  \\ \hline
{Material for the primary and secondary mirrors} & A lithium aluminosilicate glassceramic \\ & material with thermal expansion coefficient \\ & lower than 0.05x10-6 /K between 20\deg C and 300\deg C \\ \hline
Main mirror optical diameter  & 500 mm \\ \hline
Main mirror diameter & 510 mm \\ \hline
Main mirror thickness & 60 mm \\ \hline
Secondary mirror optical diameter & 181 mm \\ \hline
Secondary mirror diameter & 186 mm \\ \hline
Secondary mirror thickness & 35 mm \\ \hline
Mechanical structure material & Aluminium and carbon fiber \\ \hline
  \end{tabular}
  }
 \end{center}

\end{table}

 \acknowledgements
We are very grateful to Jack Harvey,  Alexei A. Pevtsov,  Valentin M. Pillet,  Sanjay Gosain, Frank Hill, and Luca Bertello for the discussions/suggestions on the design of the proposed instrument. This work is partially supported by the National Institute for Space Research (INPE/Brazil). 
J.~P.~acknowledges funding from IAU to attend IAUS305 and UAH-travel grants. She also acknowledges projects AYA2013-47735-P and the JPI mobility grants program of Banco Santander. L.~B.~acknowledges FAPESP 2013/03085-7.  L.E.A.V acknowledges funding from INPE's Institutional Program (Projeto Institucional PCI-INPE) under the grant agreement no. 170115/2014-3 and 170177/2014-9. L.A.S also acknowledges the funding from the PCI-INPE (grant no. 313698/2014-7). A. D. L. acknowledges project CNPq 305351/2011-7.

\end{document}